\documentclass[fleqn,usenatbib]{mnras}
\usepackage[normalem]{ulem}

\DeclareRobustCommand{\VAN}[3]{#2}
\let\VANthebibliography\thebibliography
\def\thebibliography{\DeclareRobustCommand{\VAN}[3]{##3}\VANthebibliography}

\usepackage{pdflscape}
\usepackage{multicol}
\usepackage{afterpage}
\usepackage{graphicx}	
\usepackage{amsmath}	
\usepackage{amssymb}	
\usepackage{xcolor}

\usepackage{newtxtext,newtxmath}

\title[Magnetic CP stars in Orion OB1]{Spectropolarimetry of Magnetic Chemically Peculiar Stars in the Orion OB1 Association}

\author[E. Semenko et al.]{
Eugene Semenko,$^{1}$\thanks{E-mail: eugene@narit.or.th}
Iosif Romanyuk,$^{2}$\thanks{E-mail: roman@sao.ru}
Ilya Yakunin$^{2, 3}$,
Dmitry Kudryavtsev$^{2}$,
\and Anastasiya Moiseeva$^{2}$
\\
$^{1}$National Astronomical Research Institute of Thailand, Mae Rim, Chiang Mai 50180, Thailand\\
$^{2}$Special Astrophysical Observatory of the Russian Academy of Sciences, Nizhny Arkhyz 369167, Russia\\
$^{3}$Saint Petersburg State University, Saint Petersburg 199034, Russia
}

\date{Accepted XXX. Received YYY; in original form ZZZ}

\pubyear{2022}

\begin{document}
\label{firstpage}
\pagerange{\pageref{firstpage}--\pageref{lastpage}}
\maketitle

\begin{abstract}
We summarise the results of a spectropolarimetric survey of 56 chemically peculiar (CP) stars in the association of Orion OB1. We uniformly collected the observational material with the 6-m telescope BTA of the Special Astrophysical Observatory in 2013\:--\:2021. We identify 14 new magnetic CP stars with a longitudinal magnetic field exceeding approximately 500\,G. The studied sample contains 31 magnetic stars or 55\% of the whole CP population in Orion OB1. We show that the percentage of the magnetic CP stars and the field strength drops sharply with age. The mean longitudinal magnetic field in the young subgroup OB1b ($\log t=6.23$) is confidently almost three times stronger than in the older subgroups OB1a ($\log t=7.05$) and OB1c ($\log t=6.66$). In the Orion Nebula, a place with the youngest stellar population ($\log t < 6.0$), we detect the magnetic field only in 20\% of CP stars. Such occurrence drastically differs from 83\% of magnetic CP stars in the nearby subgroup OB1c. We consider this effect an observational bias caused by a significant portion of a very young population with the signatures of Herbig Ae/Be stars. The technique we used for magnetic measurements, and the quality of available data do not allow us to detect weak fields in the case of stars with a limited number of lines and emissions in spectra.
\end{abstract}




\section{Introduction}

A remarkable fraction of B, A, and F stars shows the abnormal abundances of certain chemical elements in their atmospheres. The recent catalogue of chemically peculiar (CP) stars contains 8205 records of anomalies~\citep{2009A&A...498..961R}, while the number of CP stars is growing slowly as the new dedicated surveys are carried out. The fraction of CP stars may be as high as 10-15\% of the whole population of B to F stars \citep{2017A&A...599A..66S}. Spectroscopy remains the primary, direct tool for detecting new peculiar stars \citep{2018A&A...619A..98H}. However, in the age of global photometric surveys, space-based photometry more and more often serves as a powerful instrument for the same purposes~\citep[e.g.,][]{2018MNRAS.478.2777B, 2019MNRAS.487..304D, 2021MNRAS.504.4841D}. 

A mechanism responsible for the chemical anomalies in early-type stars is a diffusion of atoms in stellar interiors. \citet{1970ApJ...160..641M} was the first to describe the theory of the atomic diffusion in stars. According to this theory, the observed non-uniform vertical and horizontal distribution of the chemical elements results from a competition between gravitational settling and radiative acceleration of atoms in a stable environment. Modern statements of the theory allow modelling of the diffusion in hot Am and HgMn-type stars and give a relatively accurate approach to the observed effects~\citep{2015ads..book.....M}. The stars mentioned above are characterised by slow rotation, which is, in contrast to the ``normal'' stars of the same spectral class, known as the most common stabilisation mechanism. Another stabilising factor is a magnetic field inherent to the CP2 group of chemically peculiar stars according to the classification proposed by~\citet{1974ARA&A..12..257P}. The strong, stable and globally organised magnetic field significantly alters the structure of the stellar atmospheres and thus should impact the stellar evolution. 

To trace the effect of the magnetic field on the evolution of CP stars, one should study the observational parameters of stars with well-determined age. Unfortunately, the accurate dating of the magnetic CP stars is a big problem. The errors in the case of field stars can often exceed 100\% of the absolute age \citep{2006A&A...450..777B}. When the stars belong to the stellar group with known age, the situation greatly improves. In attempts to use the recent data on distances to the CP stars and their fundamental parameters for the study of the evolutionary changes of the magnetic field strength, researchers turned their attention to the open clusters hosting known magnetic CP members. Studying the stars with masses in the interval from 2 to 9\,$M_\odot$, \citet{2006A&A...450..777B} and \cite{2007A&A...470..685L, 2008A&A...481..465L} found a slow decline of the magnetic field strength with age. While for least massive stars ($M\leqslant3\,M_\odot$), the rate of decay  agreed with an assumption about the conservation of magnetic flux, for more massive stars, this rate showed evidence of flux decay. Later, this result was verified by \cite{2016A&A...592A..84F} and \cite{2019MNRAS.490..274S}. The observed effects are consistent with the theoretical models developed by \cite{2020ApJ...900..113J}. These authors explain the destruction of the magnetic field by convective turbulence, which increases with mass due to the increased size of the subsurface convection zone.

Although the origin of magnetic fields in CP stars remains unclear, there are three main hypotheses explaining it. Depending on the time scale when the field was formed, the number of hypotheses can be reduced to two. The fossil field hypothesis suggests that CP2 stars inherited their field from the local galactic field in the place of birth~\citep{1989MNRAS.236..629M}. According to this hypothesis, as the flux remains conserved during the MS stage, slight yet measurable decay of the magnetic field strength should be observed. The alternative hypothesis invokes the mechanism of a turbulent dynamo in the convective layers of a star \citep[e.g.][]{2003ASPC..305....3M}. Within the frame of this hypothesis, the generation of the magnetic field is constantly running, and the strength of the field should correlate with rotation and turbulence rates. Some authors like \citet{2004Natur.431..819B} say that the magnetism of CP stars is the result of a combination of different mechanisms. A very popular in the last decade hypothesis of 'stellar mergers' \citep{2019Natur.574..211S} is an example of such combination. According to it, the dynamo generates the seed field during a short phase after stellar merging, and subsequently the field slowly decays in agreement with the hypothesis of fossil origin.

The magnetic field decay demonstrated in \citet{2006A&A...450..777B}, \cite{2007A&A...470..685L, 2008A&A...481..465L}, and \cite{2019MNRAS.483.3127S} is a powerful argument favouring the hypothesis that the magnetic field originates in the early stages of stellar life. Another arguments are the absence of a significant correlation between the strength of the magnetic field and stellar rotation \citep[e.g.][]{2019MNRAS.483.3127S, 2019MNRAS.490..274S} and the occurrence of CP stars with extremely strong fields among slow and moderate rotators.

\citet{2014A&A...561A.147B} extended the studies of CP stars' evolution onto the phenomenon of chemical anomalies. The authors discovered the weakening of anomalies for almost all analysed chemical elements in massive ($3\,M_\odot \leqslant M \leqslant 4\,M_\odot$) stars of their sample. Similar trends were reported by \cite{2019MNRAS.483.2300S} for the mean surface abundances in Ap stars. Both studies, in addition to the impact on stellar physics, show the importance of accurate dating for the chances of CP stars to be recognised by the low-resolution spectroscopy in massive surveys.

As of the early 2010s, researches on the evolution of CP2 stars were predominantly concentrated on the open clusters, which, with all their advantages, have a significant drawback~--- a statistically small incidence of peculiar stars in their population. To overcome this limitation, in \citeyear{2013AstBu..68..300R} we launched a massive observational programme aimed at the study of the magnetic field and chemical abundances of magnetic or potentially magnetic CP stars in the Orion OB1 association. Our choice was determined by the accurate dating of the population, relatively close distance, and the observational availability of the association from the installation site of the Russian 6-m telescope, the main instrument involved in our project.

This paper is structured as follows: in Section \ref{struct}, we briefly describe the Orion OB1 association and its CP stars. The methods of the observational data reduction and magnetic field measurements are described in Section \ref{MField}. Concluding remarks and discussion of the results are summarised in Section \ref{discus}.

\section{The Orion OB1 Association}\label{struct}
\subsection{General information } 

The Orion OB1 association with the centre distanced approximately 380\,pc~\citep{2017A&A...608A.148Z} from the Sun is one of the closest regions of ongoing star formation in the solar vicinity. The association forms a structure extended for 150\,pc in-depth and primarily located in the southern hemisphere of the Galaxy within galactic latitudes $b$ from +5 to $-25^\circ$. In the sky, the association forms contours of a constellation of Orion and covers almost 200 deg$^2$ of its area.

The vast majority of the constellation stars belong to the association, except a fraction of bright and thus less distant objects projected on the area but unrelated to the association. Besides the Main Sequence B and A stars considered in this article, Orion OB1 hosts younger Herbig Ae/Be and T Tau stars.

The Orion OB1 association has a complex structure that comprises numerous open clusters, stellar groups, and molecular clouds. The Orion Nebula (M42), the second brightest diffuse nebula visible to the naked eyes, with an ongoing star formation, is also an essential part of the association.

\begin{figure*}
\centering\includegraphics[width=0.9\textwidth]{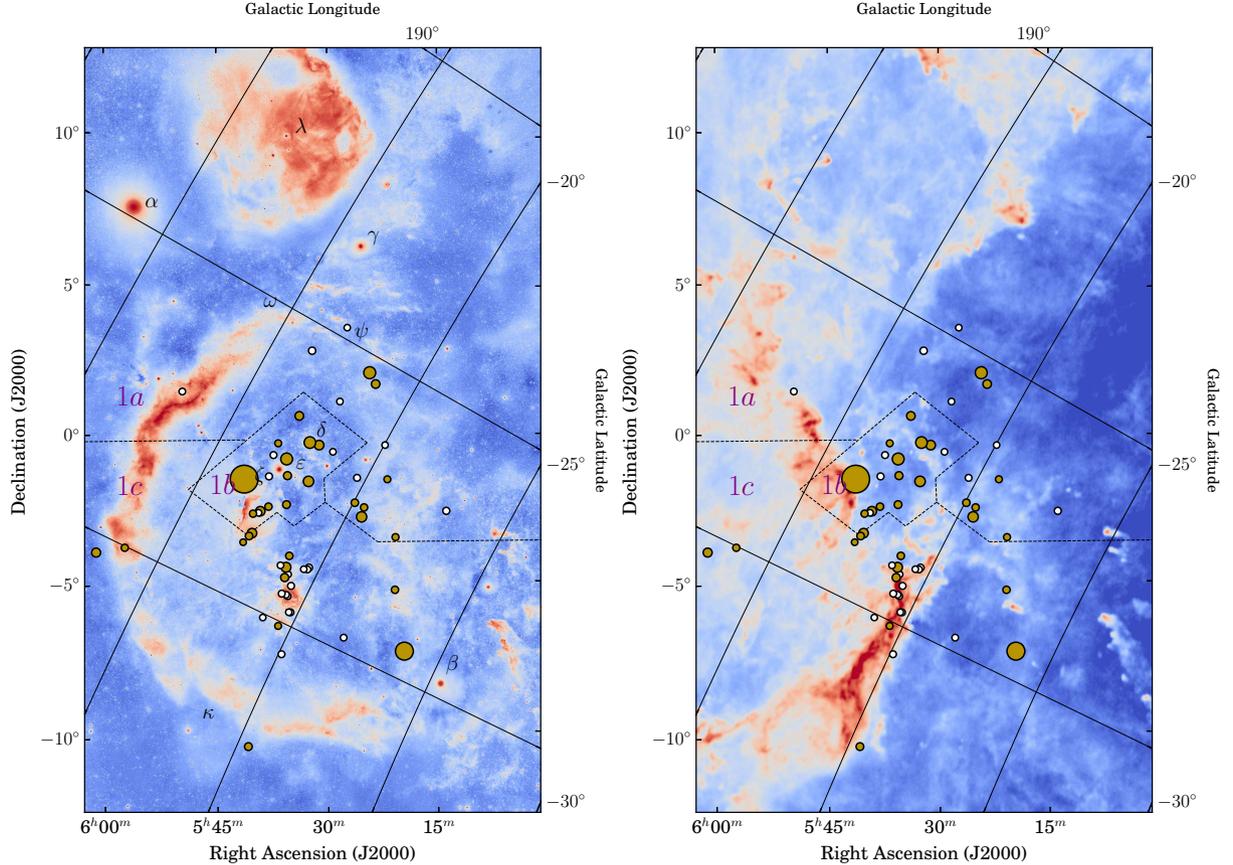}
\caption{General scheme of the association of Orion OB1 with the chemically peculiar stars, marked by colour symbols depending on their magnetic field. The non-magnetic CP stars ($\chi^{2}/\nu<5$, Sec. \ref{sec:mfield}) are marked by empty circles. Stars with a measurable magnetic field are indicated by filled circles with their size proportional to the field strength. The left panel is a false-colour image of Orion in visible light with pronounced gas complexes (Original photo by Rogelio Bernal Andreo). The right panel shows the dust emission in Planck data, taken at 353, 545 and 857 GHz. Credit: ESA and the Planck Collaboration.}
\label{fig:scheme}
\end{figure*}

Relying on the knowledge about spatial motion and evolutionary status of the stars in Orion, \citet{1964ARA&A...2..213B} set apart four subgroups with slightly varying ages and denoted them with letters from a to d. In this scheme, subgroup 1a forms the upper part of the constellation, 1b is a narrow band surrounding the Orion Belt, 1c is a lower part of the constellation with an embedded subgroup 1d corresponding to the region of the Orion Nebula (Fig. \ref{fig:scheme}). The boundaries of the association and its division were subjected to revision after the appearance of the new information about the spatial properties of the stellar population. For example, from the analysis of HIPPARCOS data, \citet{2014A&A...564A..29B} found the presence of an older foreground population in the direction to the Orion Nebula Cluster~(ONC). \citet{2015A&A...584A..26B} identified a new stream called Orion X, which extends from approximately 150--200\,pc to the front edge of the subgroup 1a. The authors believe that this stream may be a part of the association as it follows the age sequence of known subgroups. The existence of these structures later was confirmed in \citet{2017A&A...608A.148Z, 2019A&A...628A.123Z} and \citet{2020A&A...643A.114C} with the use of much more accurate data obtained in the GAIA mission~\citep{2016A&A...595A...1G}. New research has shown that the subgroups introduced by \citet{1964ARA&A...2..213B} have a non-homogeneous population, and the real picture of clustering can be more complicated.

It is believed that the visible fragmentation of the population in Orion OB1 reflects the sequential evolution of the whole association. \citet{1994A&A...289..101B}, in their fundamental paper, give the following distribution of the average age: 11.4\,Myr for 1a, 1.7\,Myr for 1b, 4.6\,Myr for 1c, and less than 1\,Myr for the youngest subgroup 1d. The recent studies, based on the GAIA parallaxes and new results of spectroscopic and photometric surveys, have revealed a more intricate structure and age distribution. For example, \cite{2018AJ....156...84K}, \cite{2019A&A...628A.123Z} and \cite{2020A&A...643A.114C} found the populations of different ages disseminated within the spatially overlapping structures. The authors explain this effect by far more complex formation history than it was believed before. They agree that the star formation in Orion started around 13\:--\:15 Myr (20 Myr in \cite{2019A&A...631A.166K}) and then spread throughout the different regions of the association. In Orion A and B molecular clouds this process is still ongoing. Despite some differences in detailed classification, the age of the individual subgroups estimated in above-cited papers is in general consistent with the values published in \citet{1994A&A...289..101B}. However, we failed in attempts to identify our sample stars in the most recent studies. In the cases when the individual stars were found in published data, they had the wrong or incomplete characteristics, first of all, radial velocities and effective temperatures. At the same time, we have no good reason to expect the critical differences in the ages of the same mass stars located closely within a limited volume. Thus, for simplification, below, we follow the original publication by \citet{1994A&A...289..101B} in terms of the age and membership of the stars as it was initially adopted in our establishing paper in \citeyear{2013AstBu..68..300R}.

\subsection{Chemically peculiar stars in Orion OB1}

We launched the observational programme in \citeyear{2013AstBu..68..300R}. To create the list of known or potentially magnetic stars, we cross-correlated data from \citet{1994A&A...289..101B} with the catalogue of CP stars by \citet{2009A&A...498..961R}. From the total number of 814 stars, we selected 85 A and B-type stars with known peculiarities. Among them, 24 stars were classified as Am. Despite the recent reports about the discovery of the extremely weak magnetic field ($\sim$1~G) in Am stars like Alhena~\citep[see][for the full list of Am stars with ultra-weak fields discovered thus far]{2020MNRAS.492.5794B}, this variant of CP stars, in general, is considered non-magnetic. Moreover, drawing on the information about parallaxes and proper motion, we conclude that all selected Am-stars are located closer to the Sun than the front edge of the association. Even though, it might be an interesting task to investigate the relation of these stars to any of the streams connected to Orion OB1, we excluded all confirmed Am stars from a spectropolarimetric monitoring for this project. The subsequent observations showed that, among the remaining 61 CP stars, 3 objects resembled HgMn-type of anomalies (considered as non-magnetic in \citealt{2011A&A...525A..97M}, and \citealt{2014psce.conf..389K}) and a Be star wrongly classified as CP (also non-magnetic following \citealt{2016ASPC..506..207W}).

The complete list of identified chemically peculiar stars in Orion OB1 contains 56 objects. In Table~\ref{tab:tab1}, for these stars, we give the information about their number in the Henry Draper and \citet{2009A&A...498..961R} catalogues, type of peculiarity, projected rotational velocity and period of axial rotation, parallaxes and proper motions, multiplicity and magnetic characteristics with corresponding references. The stars are organized into four groups according to their membership and arranged in the order of increasing HD numbers.

\begingroup
\setlength{\tabcolsep}{1.5pt}
\renewcommand{\arraystretch}{1.35}
\begin{table*}
\caption{List of the chemically peculiar stars identified in Orion OB1 with information about their peculiarities, rotation ($v_\mathrm{e}\sin i$, $P_\mathrm{rot}$), distance ($\pi_\mathrm{GAIA}$ and $d$), proper motions ($\mu_{\alpha}$, $\mu_{\delta}$), binarity and root mean squared longitudinal magnetic field ($B_\mathrm{rms}$, $\sigma_\mathrm{rms}$, $\chi^{2}/\nu$) measured using the regression method (See Sec. \ref{sec:mfield}). The number of individual observations used for evaluation of $B_\mathrm{rms}$ is $n$. The source of the information is given in square brackets in the last column, unless it is specified individually in the corresponding column.}
\label{tab:tab1}
\begin{tabular}{lccccccclccccl}
\hline
HD  & Renson & Sp, pec & $v_\mathrm{e}\sin i$ &  $\pi_\mathrm{GAIA}$ & $\mu_{\alpha}$ & $\mu_{\delta}$      &  $d$      &  Binarity        &  $P_\mathrm{rot}$     &  $B_\mathrm{rms}$    & $\sigma_\mathrm{rms}$    & $\chi^{2}/\nu (n)$ &    Reference \\
    &        &         & (km\,s$^{-1}$) & (mas)    & (mas\,yr$^{-1}$)  & (mas\,yr$^{-1}$)   & (pc)        &                      & (days) &    (G)   &    (G)    &         &     \\
\hline
\multicolumn{12}{c}{Subgroup 1a} \\
33917      &  8560   &  A0 Si        & 140       &   2.656 & 0.206 & 0.120 & $376.5^{+3.6}_{-3.5}$   &  var RV [42]    &  1.849 [51]    &    332  &  195  & 1.9 (4)        &  [37]    \\
34859$^*$  &  8900   &  A0 Si        &  90       &   3.181 & 0.782 & -0.436 & $314.3^{+2.0}_{-2.0}$   &  no data        &  1.046         &    303  &  120  & 9.9 (4)       &  [37]    \\
35008$^*$  &  8940   &  B8 Si        & 210 [37]  &   5.015 & -2.667 & 0.472 & $199.4^{+1.6}_{-1.5}$   &  var RV [27]    &   0.462 [51]   &    258  &  155  & 7.0 (4)    &  [3, 37] \\
35039      &  8953   &  B2 He-r      & 5 [49]    &   2.859 & -0.077 & 1.563 & $349.8^{+20.5}_{-18.3}$ &  SB1 [1]        &   2.8 (:) [51] &     75  &   49  & 2.3 (4)    &  [37]    \\
35177$^*$  &  8980   &  B9 Si        & 200       &   2.879 & 2.671 & 0.614 & $347.4^{+5.4}_{-5.2}$   &  no data        &   0.528 [28]   &    940  &  275  & 12.4 (5)  &  [37]     \\
35298$^*$  &  9020   &  B6 He-wk     & 55 [36]   &   2.789 & 1.214 & 0.186 & $358.6^{+7.5}_{-7.2}$   &  var RV [36]    &   1.855 [28]   &   2330  &  330  & 71.9 (15)   &  [11, 50] \\
35456$^*$  &  9070   &  B7 He-wk     & 22 [16]   &   2.887 & -4.807 & -1.410 & $346.4^{+8.4}_{-8.0}$   &  double [6]     &   4.982 [51]   &   440   &   36  & 37.7 (6)  &  [11, 23, 34, 37] \\
35502$^*$  &  9120   &  B6 SrCrSi    & 75 [46]   &   2.749 & -0.533 & -1.952 & $363.7^{+6.3}_{-6.1}$   &  SB3 [46]       & 0.8538 [46]    &   1595  &  301  & 31.0  (25) &  [11, 23, 45] \\
35575      &  9160   &  B3 He-wk     & 150       &   2.905 & 1.094 & 0.688 & $344.2^{+8.3}_{-7.9}$   &  no data        & 0.9852 [51]    &    220  &  196  & 1.2 (4)   &  [37]      \\
35730      &  9230   &  B4 He-wk     &  80       &   3.042 & 0.824 & -0.263 & $328.8^{+5.3}_{-5.1}$   &  no data        & 1.182 (:) [51] &     90  &  156  & 0.5 (5)    &  [37]      \\
35881      &  9250   &  B8 He-wk     & 340 [51]  &   2.954 & 1.553 & -0.121 & $338.6^{+4.8}_{-4.6}$   &  no data        & 0.870 (:) [51] &    130  &  164  & 0.8 (6)   &  [34, 37]  \\
36429      &  9420   &  B6 He-wk     &  75       &   2.958 & 1.889 & 0.409 & $338.0^{+5.7}_{-5.6}$   &  double         & 0.73 (:) [51]  &    518  &  375  & 2.28 (5) &  [6, 11, 14]   \\
38912      & 10450   &  B8 Si        & 100       &   2.253 & -0.452 & -5.243 & $443.9^{+3.8}_{-3.7}$   &  no data        & 1.463 [28]     &    165  &  163  &  1.2 (3)  &  [37]       \\
294046$^*$ &  9190   &  B9 Si        & 120 [51]  &   2.826 & 0.929 & -1.551 & $353.8^{+3.9}_{-3.8}$   &  no data        & 0.8985         &    496  &  164  &  13.3 (4)   &  [37]       \\
\multicolumn{12}{c}{Subgroup 1b} \\
36046      &  9290   &  B8 He-wk     & 100 [38]   &   2.889 & 0.260 & -2.595 & $346.2^{+6.2}_{-6.0}$    &  var RV [38]  & 1.036 [51]     &    119  &  253  & 0.4 (5)  &  [4, 38]  \\
36313$^*$  &  9370   &  B8 He-wk     & 160 [38]   &   2.593 & 1.991 & 5.639 & $385.7^{+105.9}_{-68.4}$ &  SB2 [38]     &  0.5891 [38]   &   1337  &  500  & 7.2 (12)  &  [4, 11, 34, 38] \\
36485$^*$  &  9440   &  B2 He-r      &  32 [24]   &   2.624 & 1.523 & -1.653 & $381.0^{+8.0}_{-7.7}$    &  SB2 [24]     & 1.47774 [24]   &   2156  &  132  & 635.0 (14) &  [8, 9, 38] \\
36526$^*$  &  9460   &  B8 He-wk,Si  &  50 [38]   &   2.387 & 0.465 & -1.748 & $419.0^{+7.4}_{-7.1}$    &  double [38]  & 1.541 [38]     &   1695  &  137  & 539.3 (6)  &  [11, 38]    \\
36668$^*$  &  9560   &  B8 He-wk, Si &  60 [38]   &   2.546 & -0.281 & -1.078 & $392.9^{+7.4}_{-7.2}$    &  no data      &   2.121 [38]   &    953  &  105  & 203.5 (8) &  [11, 35, 38] \\
36955$^*$  &  9740   &  A2 CrEu      &  26 [38]   &   2.235 & -0.120 & 0.274 & $447.5^{+9.1}_{-8.7}$    &  double       &   2.2835 [38]  &    708  &   90  &  93.9 (6) &  [20, 38]   \\
37140$^*$  &  9910   &  B8 SiSr      &  30 [38]   &   2.383 & -1.231 & 1.950 & $419.7^{+8.7}_{-8.4}$    &  triple [32]  &   2.704 [38]   &    270  &  107  &  9.6 (6)  &  [11, 38]  \\
37235      &  9960   &  B9 He-wk     & 320        &   2.611 & -0.910 & -0.537 & $383.1^{+6.6}_{-6.4}$    &  no data      &   0.4846       &    227  &  170  &  2.3 (5)  &  [38]       \\
37321      & 10000   &  B5 He-wk     & 130        &   2.915 & 1.313 & -0.572 & $343.0^{+6.9}_{-6.6}$    &  double       & 0.563 (:) [51] &    290  &  202  &  1.4 (5)  &  [38]       \\
37333$^*$  & 10010   &  Ap Si        &  50 [38]   &   2.982 & -2.686 & -3.494 & $335.3^{+4.2}_{-4.1}$    &  no data      &   1.6833 [38]  &    433  &  120  &  14.6 (6) &  [4, 38]    \\
37479$^*$  & 10080   &  B2 He-r      & 150 [38]   &   2.308 & 1.284 & -0.205 & $433.3^{+12.5}_{-11.8}$  &  SB           &   1.190 [13]   &   1307  &  324  &  46.6 (12)  &  [9, 22, 38]    \\
37525      & 10110   &  B6 He-wk     & 160 [38]   &   2.513 & 1.317 & -0.693 & $398.0^{+7.7}_{-7.4}$    &  double       & 4.9863 (:) [51]&    127  &  192  &  0.3 (5)  &  [38, 48]    \\
37633$^*$  & 10130   &  B9 EuSi      &  35 [38]   &   2.401 & -3.599 & 2.432 & $416.6^{+6.6}_{-6.4}$    &  no data      &   1.573 [28]   &    382  &   97  &  25.5  (6) &  [4, 38]    \\
37776$^*$  & 10190   &  B2 He-r      &  91 [19]   &   2.546 & 2.743 & 1.671 & $392.7^{+8.3}_{-7.9}$    &  single       &  1.5387 [26]   &   8644  &  500  &  298.5 (30)  &  [10, 18, 19, 33, 38, 47] \\
290665$^*$ &  9700   &  B9 SrCrEuSi  &  20 [51]   &   2.548 & -2.033 & -0.759 & $392.4^{+3.3}_{-3.3}$    &  no data      &   5.1644 [51]  &   2260  &   54  &  2343.4 (10)  &  [4, 20, 38] \\
\multicolumn{12}{c}{Subgroup 1c} \\
34736$^*$  &  8860   &  B9 Si        & 73 [41]    &   2.685 & -0.300 & -0.234 & $372.4^{+7.7}_{-7.4}$    &  SB2 [41]     &  1.28 [39]     &   4700  &  350  & $>1000$ (130)    &  [39, 41] \\
34889$^*$  &  8929   &  B9 Si        & $<20$ [39] &   2.836 & 0.300 & -2.575 & $352.6^{+3.9}_{-3.8}$    &  no data      &  3.631 [39]    &    433  &   91  & 50.5 (6)      &  [37, 39]  \\
35901      &  9255   &  B9 Si        &  65        &   1.522 & -0.921 & -2.624 & $656.8^{+11.7}_{-11.3}$  &  no data      &  2.949         &    90   &   60  & 2.8 (7)   &  [39]      \\
36540      &  9480   &  B7 He-wk     &  75 [39]   &   2.418 & 0.589 & -0.021 & $413.6^{+4.6}_{-4.5}$    &  double       &  2.172 [39]    &    258  &  120  & 3.8 (4)  &  [4, 11, 35, 39] \\
36559      &  9500   &  A0 p         & 140        &   3.065 & 1.170 & -0.246 & $326.2^{+3.3}_{-3.2}$    &  no data      &   $-$          &    180  &  200  & 0.7 (4)  &  [39]      \\
36629      &  9550   &  B3 He-wk     &   5 [17]   &   2.331 & 0.853 & -2.568 & $429.0^{+5.8}_{-5.6}$    &  no data      &   15.98 [30]   &    94   &   69  & 2.4 (7) &  [4, 39, 48] \\
36899      &  9690   &  B9 Sr        & 250        &   2.539 & 1.197 & -0.150 & $393.8^{+4.1}_{-4.0}$    &  var RV       &   $-$          &    267  &  164  &  3.0 (7)  &  [39]      \\
36916$^*$  &  9700   &  B8 He-wk,Si  & 55 [39]    &   3.852 &-6.256 & -1.425 & $259.6^{+3.2}_{-3.2}$    &  var RV [39]  &   1.565 [39]   &    461  &  181  &  7.2 (7) &  [12, 35, 39] \\
\hline
\end{tabular}
\end{table*}

\begin{table*}
\contcaption{List of the chemically peculiar stars identified in Orion OB1 with information about their peculiarities, rotation ($v_\mathrm{e}\sin i$, $P_\mathrm{rot}$), distance ($\pi_\mathrm{GAIA}$ and $d$), proper motions ($\mu_{\alpha}$, $\mu_{\delta}$), binarity and root mean squared longitudinal magnetic field ($B_\mathrm{rms}$, $\sigma_\mathrm{rms}$, $\chi^{2}/\nu$) measured using the regression method (See Sec. \ref{sec:mfield}). The number of individual observations used for evaluation of $B_\mathrm{rms}$ is $n$. The source of the information is given in square brackets in the last column, unless it is specified individually in the corresponding column.}
\begin{tabular}{lccccccclccccl}
\hline
HD  & Renson & Sp, pec & $v_\mathrm{e}\sin i$ &  $\pi_\mathrm{GAIA}$ & $\mu_{\alpha}$ & $\mu_{\delta}$      &  $d$      &  Binarity        &  $P_\mathrm{rot}$     &  $B_\mathrm{rms}$    & $\sigma_\mathrm{rms}$    & $\chi^{2}/\nu (n)$ &    Reference \\
    &        &         & (km\,s$^{-1}$) & (mas)    & (mas\,yr$^{-1}$)  & (mas\,yr$^{-1}$)   & (pc)        &                      & (days) &    (G)   &    (G)    &         &     \\
\hline
36918      &  9710   &  B9 He-wk     &  75 [39]   &   2.665 & 1.205 & 0.508 & $375.3^{+5.0}_{-4.8}$    &  triple [39]    & 0.892 [51]   &    234  &  167  &  1.7 (5)  &  [4, 39]   \\
36958      &  9750   &  B3 He-wk     &  $<20$     &   2.755 & 0.971 & 0.113 & $363.0^{+10.4}_{-9.9}$   &  var RV [39]    & 37.13 [30]  &     86  &  115  &  1.1 (7) &  [39]      \\
36960      &  9780   &  B0 Si        &  45 [39]   &   2.617 & 1.111 & 1.681 & $382.2^{+18.4}_{-16.8}$  &  var RV        & 0.995 (:) [51] &     93  &   63  &  1.1 (4)  &  [4, 39]   \\
36997$^*$  &  9810   &  B9 SiSr      &  30        &   2.360 & -0.796 & 1.694 & $423.7^{+12.3}_{-11.6}$  &  SB2           & 6.00 [7]   &    716  &   40  &  400.0 (10)  &  [39]      \\
37017$^*$  &  9820   &  B2 He-r      & 100 [39]   &   2.784 & 1.477 & 1.554 & $359.3^{+9.0}_{-8.5}$    &  SB2           &  0.901 [39]  &   1460  &  280  &  33.6 (25) &  [9, 21, 39] \\
37058$^*$  &  9850   &  B3 He-wk,Sr  & 25 [39]    &   2.618 & 1.490 & 0.742 & $382.0^{+6.5}_{-6.3}$    &  no data       & 14.612 [39]  &    775  &   75  &  186.5 (7)  &  [4, 12, 25, 39] \\
37129      &  9890   &  B3 He-wk     & 54 [39]    &   2.636 & 1.193 & -0.465 & $379.3^{+6.7}_{-6.4}$    &  var RV [39]    &   $-$       &    128  &  130  &  1.0 (5)  &  [39]      \\
37151      &  9930   &  B8 He-wk     & 30 [39]    &   4.335 & -3.643 & 5.559 & $230.7^{+2.4}_{-2.4}$    &  var RV [39]    & 5.6732 [29] &    232  &  118  &  2.1 (5)  &  [5, 11, 39] \\
37210$^*$  &  9950   &  B8 He-wk,Si  & $<20$ [39] &   2.102 & 1.185 & 0.729 & $475.7^{+11.5}_{-11.0}$  &  var RV [39]    & 11.0494 [15] &    232  &   80  &  10.1 (4) &  [4, 11, 39] \\
37470      & 10070   &  B8 Si        & 130 [39]   &   2.410 & 4.422 & -2.318 & $414.9^{+9.3}_{-8.9}$    & SB2 [39]        & 0.6148 [39]   &     71  &  160  &  0.3 (6) &  [4, 39]    \\
37642$^*$  & 10150   &  B9 He-wk,Si  & 85 [39]    &   2.712 & 0.871 & -0.834 & $368.7^{+4.8}_{-4.7}$    &  var RV [39]    & 1.07977 [29] &   1311  &  234  &  27.6 (10)  &  [11, 20, 39]  \\
37687$^*$  & 10160   &  B7 He-wk,Si  & $<20$ [39] &   2.244 & -6.717 & -1.132 & $445.6^{+18.3}_{-16.9}$  &  var RV [39]    & 3.85238 [30] &    525  &   31  &  752.0 (6)  &  [2, 39]   \\
37807$^*$  & 10200   &  B4 He-wk     &  20 [39]   &   2.436 & 2.774 & -2.894 & $410.5^{+7.7}_{-7.4}$    &  no data       &  0.532 [39]   &    144  &   73  &  5.3 (5) &  [39]       \\
37808$^*$  & 10210   &  B9 Si        &  25 [39]   &   5.405 & -0.681 & -0.684 & $185.0^{+1.8}_{-1.8}$    &  no data       & 1.0991 [15]  &    615  &  103  &  39.6  (5)  &  [39]       \\
40146$^*$  & 10710   &  A0 Si        &  40        &   2.115 & 6.501 & 0.924 & $472.9^{+4.7}_{-4.6}$    &  no data       & 3.5604       &    374  &   56  &  48.0 (3)  &  [39]       \\
40759$^*$  & 10900   &  A0 CrEu      &  26 [51]   &   2.434 & -2.990 & 3.682 & $410.9^{+8.7}_{-8.3}$    &  SB3 [51]       & 3.37484 [51]  &   1120  &   65  &  418.4 (14) &  [39]       \\
\multicolumn{12}{c}{Subgroup 1d} \\
36982      &  9800   &  B2 He-r      &  80 [39]   &   2.449 & 1.619 & 1.782 & $408.3^{+4.0}_{-4.0}$    &  no data       & 0.744 (:) [51] &    104  &   75  &  1.1 (7) &  [4, 31, 39, 40, 43, 44] \\
37041      &  9830   &  B0 He-r      & 120 [39]   &   2.973 & 1.089 & 2.394 & $336.4^{+26.0}_{-22.5}$  &  SB2           &   $-$    &    107  &  148  &  2.6 (4)  &  [4, 39]   \\
37114      &  9880   &  B9 p         & 120 [39]   &   3.031 & 0.662 & -1.694 & $330.0^{+3.1}_{-3.1}$    &  var RV [39]    &   $-$    &    248  &  111  &  3.9 (4)  &  [39]      \\
\hline
\end{tabular}
[1]  \cite{2005ApJ...629..507A}, [2]  \cite{2007A&A...475.1053A}, [3]  \cite{2006A&A...450..777B}, [4]  \cite{2006A&A...450..777B}, [5]  \cite{2015A&A...583A.115B}, [6]  \cite{2012AstBu..67...44B}, [7]  \cite{2015A&A...581A.138B}, [8]  \cite{1989ApJ...346..459B}, [9]  \cite{1987ApJ...323..325B}, [10]  \cite{1979ApJ...228..809B}, [11]  \cite{1981ApJ...249L..39B}, [12]  \cite{1983ApJS...53..151B}, [13]  \cite{2020A&A...639A..81B}, [14] \cite{2009MNRAS.394.1338B}, [15]  \cite{1998A&AS..127..421C}, [16]  \cite{2016AstBu..71..489C}, [17]  \cite{2005ESASP.560..571G}, [18]  \cite{2000AstL...26..177K}, [19]  \cite{2011ApJ...726...24K}, [20]  \cite{2006MNRAS.372.1804K}, [21]  \cite{1978ApJ...224L...5L}, [22]  \cite{1978ApJ...224L...5L}, [23]  \cite{2007A&A...470..685L}, [24]  \cite{2010MNRAS.401.2739L}, [25]  \cite{1997A&AS..124..475M}, [26]  \cite{2011A&A...534L...5M}, [27]  \cite{1991ApJS...75..965M}, [28]  \cite{2017MNRAS.468.2745N}, [29]  \cite{1984A&AS...55..259N}, [30]  \cite{2018AJ....155...39O}, [31]  \cite{2008MNRAS.387L..23P}, [32]  \cite{2014AstBu..69..296R}, [33]  \cite{1998BSAO...45...93R}, [34]  \cite{2016AstBu..71..436R}, [35]  \cite{2017AstBu..72..165R}, [36]  \cite{2018AstBu..73..178R}, [37]  \cite{2019AstBu..74...55R}, [38]  \cite{2021AstBu..76...39R}, [39]  \cite{2021AstBu..76..163R}, [40]  \cite{2017A&A...599A..66S}, [41]  \cite{2014AstBu..69..191S}, [42]  \cite{2019ASPC..518...31S}, [43]  \cite{2018MNRAS.475.5144S}, [44]  \cite{2019MNRAS.490..274S}, [45]  \cite{2016MNRAS.460.1811S}, [46]  \cite{2016MNRAS.460.1811S}, [47]  \cite{1985ApJ...289L...9T}, [48]  \cite{2016MNRAS.456....2W}, [49]  \cite{2007AJ....133.1092W}, [50]  \cite{2013AstBu..68..214Y}, [51] This work
\end{table*}
\endgroup

\begin{figure}
\centering\includegraphics[width=0.9\columnwidth]{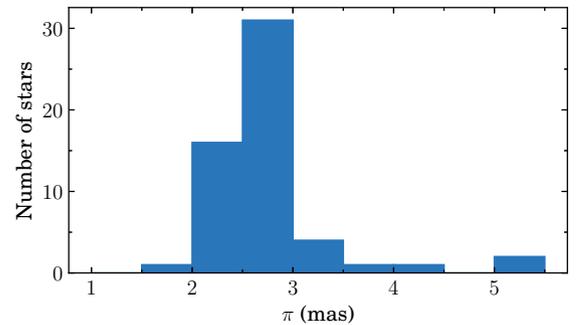}
\caption{Distribution of GAIA EDR3 parallaxes for the stars from Table~\ref{tab:tab1}.}
\label{fig:plx}
\end{figure}

\begin{figure}
\centering\includegraphics[width=0.9\columnwidth]{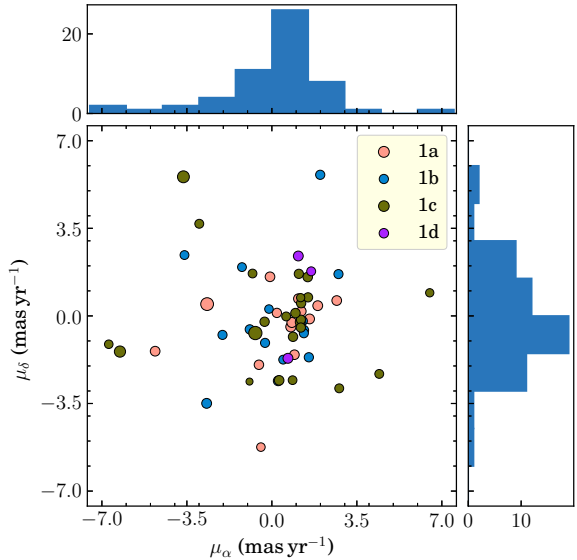}
\caption{Proper motion diagram for the stars from Table~\ref{tab:tab1}. Marker size is proportional to parallax. Histograms show the corresponding distributions for the whole sample of the stars.}
\label{fig:pm}
\end{figure}

The membership verification of the Orion stars is a complex problem. Due to the predominant systemic motion of the association away from the Sun, the measured parallaxes and proper motions of Orion stars are small and often inaccurate. At the beginning of the project, during the preselection phase, the HIPPARCOS data in \cite{2007A&A...474..653V} reduction were the only reliable source of information about distances to the examined stars. We cross-checked the astrometric distances with the values expected from spectroscopic classification. This helped us to establish the membership of several CP stars which appeared initially to be far beyond the association. Now, after the publication of the GAIA EDR3 \citep{2021A&A...649A...1G}, we can see that our approach is correct and most of the selected stars are the true members of Orion OB1. The mean values and distribution of parallaxes (Fig.~\ref{fig:plx}) as well as the proper motions (Fig. \ref{fig:pm}) certify this conclusion.

Together with astrometry, stellar radial velocity is an important criterion of the member selection. In our work, however, we faced a serious problem. Whereas the brightest CP stars in Orion had scarcely measured radial velocities, many of them later were found in binary and multiple systems and showed variable radial velocity \citep{2019ASPC..518...31S}. Stellar binarity can also be responsible for the presence of outliers with high parallaxes in our sample.

We consider our sample of CP stars complete or close to complete. This conclusion is based on the following reasoning. The average distance to the association is 390\:--\:400 pc~\citep{2020NewAR..9001549W}, which corresponds to a distance modulus of about 8 mag. The most probable objects with chemical peculiarities in Orion OB1 are Bp stars with an absolute magnitude $M$ ranging from $-3$ to $0$ mag. At the same time, in Orion, it was identified only a few early Ap stars with $M=+1^{\rm m}$. Even with the circumstellar and interstellar extinction taken into account, all such stars must appear brighter than 10 mag in the $V$-band. Bright stars like these are mostly well-studied by low-resolution spectroscopy or photometrically, listed in different catalogues, and have little risk of being omitted.

\subsection{CP stars of the Orion Nebula}\label{onc}
In the structure of the association, the Orion Nebula has a special place. This region, rich in gas and dust, is one of the closest to the Sun where processes of massive star formation are undergoing. In Blaauw's classification, the core of the nebula is recognised as an individual subgroup 1d. However, careful examination of the stellar population in subgroup 1c shows that some of its stars also belong to the nebula. Subgroup 1d, in this respect, can be considered as a younger subsystem of a bigger subgroup 1c.

The stars located in the Orion Nebula were catalogued by \cite{1954TrSht..25....1P}. Using his list as the reference, we supplemented it with the up-to-date parallaxes, interstellar reddening and polarisation, and, in subgroups 1c and 1d, eventually selected 16 CP stars with the highest probability of membership in the nebula. The full list of such stars is given with all auxiliary information in Table \ref{tab:insideON}. In the columns, for each star, we give its number in the Henry Draper and Parenago catalogues, distance $d$ computed from the GAIA parallaxes, interstellar polarisation $P$ \citep{2000AJ....119..923H}, total extinction $A_\mathrm{V}$ \citep{1994A&A...289..101B}, and magnetic field characteristics (from Tab. \ref{tab:tab1}). In Table \ref{tab:insideON}, by an asterisk, we mark the star HD\,36916, which is present in Parenago's list but its location is obviously closer than the front edge of the nebula. We count this star with the rest of non-members stars listed in Table \ref{tab:outsideON}.

\begin{table}
\setlength{\tabcolsep}{3.7pt}
\caption{Stars from subgroups 1c and 1d located inside the Orion Nebula. Columns contain star's number in the HD and Parenago catalogues, distance $d$, interstellar polarisation $P$, total extinction $A_\mathrm{V}$, and characteristics of the magnetic field $B_\mathrm{rms}$, $\sigma_\mathrm{rms}$, $\chi^2/n$ obtained by the regression method.}
\label{tab:insideON}
\begin{center}
\begin{tabular}{lccccccl}
\hline
HD     &   Parenago     &   $d$   &  $P$ &  $A_\mathrm{v}$ &  $B_\mathrm{rms}$  &  $\sigma_\mathrm{rms}$  &  $\chi^2/n$    \\
       &                & (pc)   &  (\%)   & (mag)  & (G) & (G) & \\
\hline
36540   &     867      &      414     &     1.64   &   0.59  &   258  &  120  &    3.8  \\
36559   &     908      &      326     &     0.28   &   0.05  &   180  &  200  &    0.7  \\
36629   &    1044      &     429     &     1.84   &   0.69  &    94  &  69   &    2.4  \\
36899   &    1562      &     394     &     0.89   &   0.03  &   267  &  164  &    3.0  \\
36916$^{*}$   &    1628      &  260    &     0.27   &   0.01  &   461  &  181  &    7.2  \\
36918   &    1634      &      375     &     $-$    &   0.09  &   234  &  167  &    1.7  \\
36958   &    1708      &      363     &     0.93   &   0.28  &    86  &  115  &    1.1  \\
36960   &    1728      &      382     &     0.11   &   0.07  &    93  &  63   &    1.1  \\
37017   &    1933      &      359     &     0.25   &   0.49  &  1460  &  280  &   33.6  \\
37058   &    2083      &      382     &     0.54   &   0.15  &   775  &  75   &  186.5  \\
37129   &    2314      &      379     &     0.32   &   0.12  &   128  &  130  &    1.0  \\
37210   &    2410      &      476     &     0.11   &   0.05  &   232  &  80   &   10.1  \\
37470   &    2699      &      415     &     1.51   &   0.48  &    71  &  160  &    0.3  \\
36982   &    1772      &      408     &     1.01   &   0.94  &    104  &  75  &    1.1  \\
37041   &    1993      &      336     &     0.79   &   0.62  &    107  &  148  &    2.6  \\
37114   &    2284      &      330     &     0.39   &   0.04  &   248  &  111  &    3.9  \\
\hline
\end{tabular}
\end{center}
\end{table}

\begin{table}
\caption{Stars of subgroup 1c located outside of the Orion Nebula. The meaning of the columns is the same as in Table \ref{tab:insideON}.}
\label{tab:outsideON}
\begin{center}
\begin{tabular}{lcccccl}
\hline
HD    &   $d$ &  $P$ &  $A_\mathrm{v}$ &  $B_\mathrm{rms}$  &  $\sigma_\mathrm{rms}$  &  $\chi^2/n$    \\
      &  (pc)  & (\%)  & (mag)  & (G)  & (G)  &  \\
\hline
34736   &    372    &    0.18   &    -     &   4700   &   350   &  $>1000$ \\
34889   &    353    &    0.28   &   0.01   &    433   &   91    &    50.5 \\
35901   &    657    &    0.62   &   0.16   &    90   &   60    &     2.8 \\
36997   &    424    &     $-$   &   0.26   &    716   &   40    &    400.0 \\
37151   &    231    &    0.13   &   0.04   &    232   &   118   &     2.1 \\
37642   &    369    &    $-$    &   0.14   &   1311   &   234   &    27.6 \\
37687   &    446    &    $-$    &   0.51   &    525   &   31    &   752.0 \\
37807   &    411    &    0.30   &   0.15   &    144   &   73    &     5.3 \\
37808   &    185    &    $-$    &   0.03   &    615   &   103   &    39.6 \\
40146   &    473    &    $-$    &   0.48   &    374   &   56    &    48.0 \\
40759   &    411    &    $-$    &   0.14   &   1120   &   65    &   418.4 \\
36916$^{*}$   &    260    &    0.27   &   0.01   &    461   &   181   &     7.2 \\
\hline
\end{tabular}
\end{center}
\end{table}

\begin{figure}
\centering\includegraphics[width=0.9\columnwidth]{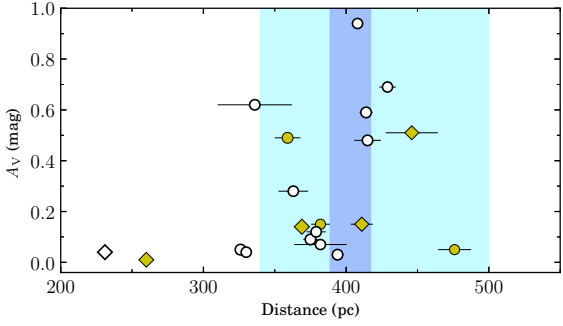}
\caption{Relation between the total extinction $A_\mathrm{V}$ and distance used as a criterion of the membership in the Orion Nebula. Circles denote the stars successfully classified as members. Magnetic stars are shown by filled symbols and their size is proportional to the field strength. Filled regions show the probable size of the nebula (light blue) and its central part (violet).}
\label{fig:criterion:distance}
\end{figure}

From the tables, one can clearly see that some non-member stars are located closer than the nebula, whose core, the Orion Nebula Cluster, or ONC, some authors place at 380\:--\:420 pc \citep[e.g.][and discussion therein]{2007MNRAS.376.1109J, 2007A&A...474..515M, 2018A&A...619A.106G} from the Sun. Small interstellar polarisation $\langle P\rangle=0.30\pm0.07\,\%$ and extinction $\langle A_\mathrm{V}\rangle = 0.175\pm0.053$ mag additionally certify the proximity of these stars. We illustrate this criterion in Fig. \ref{fig:criterion:distance}. To understand why some stars within the acceptable range of distances in Fig. \ref{fig:criterion:distance} are not recognised as members, one should remember that in the sky, these stars belonging to the subgroup 1c lay far outside the considered association region. Thus far, we estimate the average distance to the nebular CP stars as $385\pm39$ pc.

\section{Magnetic Field of CP Stars in Orion OB1}
\label{MField}

The first measurements of the stellar magnetic fields in Orion were made by \citet{1979ApJ...228..809B} with the use of a Balmer-line magnetometer. In this and subsequent works, J.~D. Landstreet and his colleagues found strong longitudinal fields in 21 CP stars in Orion OB1.

In this paper, we present the results of a spectropolarimetric survey of all 56 known and potentially magnetic CP stars from Table~\ref{tab:tab1}. The individual measurements of the longitudinal magnetic field were reported previously in \citet{2019AstBu..74...55R, 2021AstBu..76...39R, 2021AstBu..76..163R}. These papers also describe the techniques used for measuring the magnetic field. Here we describe the methods in their up-to-date state.

\subsection{Observational technique and data analysis}\label{technique}

The observations were carried out in 2013--2021 using the 6-m telescope of the Special Astrophysical Observatory in the Northern Caucasus. Spectropolarimetric data were obtained with the Main Stellar Spectrograph (MSS) of the telescope\footnote{\url{https://www.sao.ru/hq/lizm/mss/en/index.html}}~\citep{2014AstBu..69..339P}. The MSS is equipped with a dichroic polarisation analyser combined with a double image slicer and allows the simultaneous registration of two spectra with the right and left-handed circular polarisation in a single frame~\citep{2016AstBu..71..489C}. In the third order of dispersion, the MSS registers a 550\,\AA\ portion of the spectrum with an average spectral resolution $R\approx15,000$. The detector is a Teledyne E2V CCD\:42‑90 chip with $4600\times2000$ pixel format and 13.5\,$\mu$m pixel pitch.

For practical accounting for the instrumental effects of the spectropolarimeter, the analyser is equipped with a rotating quarter-wave plate that can take two positions at the angles of $-45^\circ$ and $+45^\circ$ to the optical axis. Every target is obtained twice with the different plate positions. Thus, two spectra with the opposite circular polarisation are registered by the same CCD pixels, and the instrumental shift is eliminated by simple averaging of two subsequent images.

A standard set of calibration frames required for the CCD data reduction includes bias frames, halogen lamp spectrum for flat fielding, and spectra of a ThAr hollow-cathode lamp for the wavelength calibration.

We used the context \textsc{Zeeman} written as an extension of the ESO-MIDAS framework for the processing of circularly polarised spectra obtained with the MSS~\citep{2000mfcp.proc...84K}. Processing of the CCD images includes bias and flat-field correction, subtraction of the scattered light, wavelength calibration of the spectra, and a subsequent correction to the barycentric velocity. Before further analysis, we perform a continuum normalisation of the stellar spectra.

The longitudinal magnetic field $B_\mathrm{z}$ of the survey stars can be measured due to the observational manifestation of the Zeeman effect in stellar spectra. In our studies, we usually follow two methods of magnetic field measurement, which can be classified as positional and polarisation-based. The first method is a further development of a classical technique introduced by \citet{1947ApJ...105..105B}. We derive the position of the polarised Zeeman $\sigma$-components of the spectral lines by fitting them with the Gaussian function. A distance $\Delta\lambda$ between two polarised components is proportional to the magnetic field strength $B_\mathrm{z}$: $$\Delta\lambda = 9.34\cdot10^{-13}\,\lambda_\mathrm{0}^2\,g_\mathrm{L}\,B_\mathrm{z},$$ where $\lambda_\mathrm{0}$ is a central wavelength of an unaffected line and $g_\mathrm{L}$ is the line's Land\'e factor. Such an approach works well for the spectra with sharp lines, but it can give distorted results in the case of fast axial rotation typical for Bp-stars. To verify the results of the positional method, we implemented an algorithm originally proposed in \citet{1970ApJ...160L.147A} for the magnetic field detection in the wings of hydrogen lines and extended later for general purposes in \citet{2002A&A...389..191B}. Unlike the classic method, here the longitudinal field strength depends on the measured level of circular polarisation or Stokes $V$ parameter: $$\frac{V}{I}=-4.67\cdot10^{-13}\,\lambda^2_\mathrm{0}\,g_\mathrm{L}\,\frac{1}{I}\,\frac{\mathrm{d}I}{\mathrm{d}\lambda}\,B_\mathrm{z},$$ where $I$ and $V$ denote the corresponding Stokes parameters. Our codes measure the longitudinal field from the wings of very broad lines using the adjustable spectral intervals or the entire spectrum. In this paper, we identify this method as regression.

\begin{figure}
\centering\includegraphics[width=0.96\columnwidth]{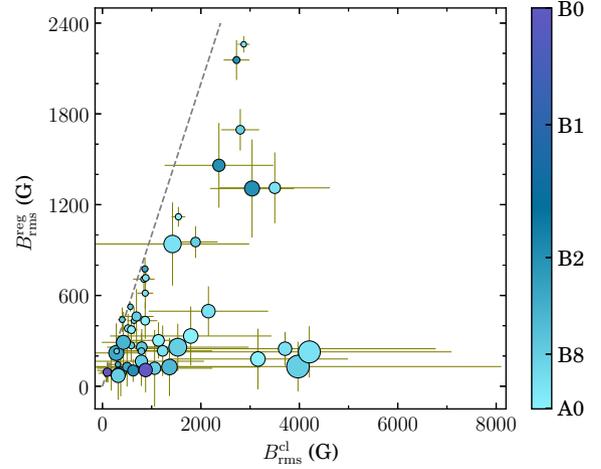}
\caption{The root-mean-square magnetic field ($B_\mathrm{rms}$) of the sample stars measured using the classic positional and regression methods. Marker size depends on a value of projected rotational velocity ($v_\mathrm{e}\sin i$) of the stars, its color corresponds to the spectral type. The individual values of $B_\mathrm{rms}^\mathrm{reg}$ are given in Table \ref{tab:tab1}, corresponding values of $B_\mathrm{rms}^\mathrm{cl}$ were published in \citet{2016AstBu..71..436R} and our series of papers \citep{2019AstBu..74...55R,2021AstBu..76...39R,2021AstBu..76..163R}.}
\label{fig:methods}
\end{figure}

The long-term application of both the methods to our research shows a good inner agreement, albeit with a few notable comments. The regression method works well in the case of very fast rotating stars, giving more precise results. However, in some exceptional cases, like an SB2 star HD\,36313 comprising hot and very fast-rotating magnetic primary, even this method fails, and the only way to detect the field is the measuring the polarised signal in the wings of hydrogen lines. Principally due to the rotational broadening, the regression method tends to give systematically lower values of the longitudinal field compared to the classic positional methods. To illustrate this effect, in Fig. \ref{fig:methods} we compare the root-mean-square magnetic field $B_\mathrm{rms}$ (see Sec. \ref{sec:mfield}) obtained for the sample stars using the two methods. We also added information about spectral types to the graph to highlight the importance of the rotation over the temperature and, consequently, the number of spectral lines.

We validate the results through the observation of the standard stars. These standards are included in the list of targets for every night and represent two different groups of objects. The first group comprises chemically peculiar stars with a well-known variation of the longitudinal field. Usually, these are the stars with accurately calculated rotational periods (e.g., 53~Cam) or those whose magnetic field, especially its polarity, can be predicted for the moment of observation, like for $\gamma$\,Equ. We recognise such stars as magnetic standards and use them to control the orientation of polarising optics of the spectrograph. The second group of standards consists of the late-type stars without detectable magnetic fields whose spectra are rich in narrow lines. Such so-called zero-field standards are used for checking the level of instrumental polarisation. Generally, the instrumental effects vanish after averaging the consecutive exposures. However, there are two related to the observational methodology issues, which potentially can lead to the incomplete accounting of the instrumental polarisation. As the MSS is a classic long-slit spectrograph, all types of short-term motion of the star projected on the slit may cause a corresponding shift in the wavelength domain. To minimise this effect, we limited the shortest exposures by 180\,s. Another problem arises from the difference in the signal-to-noise ratio (SNR) between two averaged spectra collected under unfavourable weather conditions. The continuum normalisation of the spectra before the averaging can substantially level this problem. As a result, we can keep the instrumental polarisation below the level corresponding to the longitudinal magnetic field order of 50\,G.

\subsection{Occurrence  of mCP stars and their fields}\label{sec:mfield}

We measured the longitudinal magnetic field $B_\mathrm{z}$ of all 56 stars from Table~\ref{tab:tab1} in a uniform way. Most of the sample stars have never been observed spectropolarimetrically. The sample also contains well-known magnetic stars. Our decision to include them is based on the fact of using in the past the methods (Balmer lines magnetometry, multiline spectropolarimetry, classic Babcock's technique), which are not always mutually comparable. The difference between the values of $B_\mathrm{z}$ measured from the hydrogen and metallic lines of different elements~\citep[e.g.][]{2000MNRAS.313..851W} clearly demonstrates this problem. However, our strategy gives another interesting opportunity. For some stars, our observations cover a very long timeline, which can be used to explore a long-term variation of rotational periods. In Orion, the variable rotation rate was found, for example, for HD\,37776 \citep{2008A&A...485..585M}. 

All excepting two stars in our sample were observed at least four times on different nights. In this way, we tried to catch an arbitrary phase of the rotational period, which for most of the stars under study was unknown by the moment of observation. This situation has been greatly improved with new TESS observations. In the column $P_\mathrm{rot}$ of Table \ref{tab:tab1}, to demonstrate the rotational properties of the sample, we collected periods found mostly from photometry and attributed to the stellar rotation. In this paper, we present 16 new periods found from the period analysis of the TESS data. Lightcurves extracted from the MAST archive\footnote{\url{https://dx.doi.org/10.17909/T9RP4V}} were cleaned from low-order trends before the analysis. A detailed description of the applied techniques will be given in future papers devoted to the study of individual interesting stars. At present, the listed values of $P_\mathrm{rot}$ are not final and can be improved with more advanced processing. The most uncertain values of $P_\mathrm{rot}$ in Table \ref{tab:tab1} are marked by a colon. For a few stars, the TESS periods appear doubtful because of a very complex light curve with multiple frequencies in a periodogram.

The individual measurements of the magnetic field with comments on every star were published in a series of papers \citep{2019AstBu..74...55R, 2021AstBu..76...39R, 2021AstBu..76..163R}. For these purposes, we used all the methods described in Sec.~\ref{technique}, but due to the significant intrinsic uncertainties of the classic positional method in the case of fast-rotating hot stars, in Table~\ref{tab:tab1} we operate only with data obtained using the regression method.

We compare the magnetic properties of the individual stars using the root-mean-square magnetic field $B_\mathrm{rms}$, its standard error $\sigma_\mathrm{rms}$, and test statistics $\chi^{2}/\nu$, which, following \citet{1993A&A...269..355B}, are determined as follows:

\begin{equation}
\begin{aligned}
    B_\mathrm{rms} = \langle B_\mathrm{z}^2\rangle^{1/2} = \left[\frac{1}{n}\sum_{i=1}^{n} B^2_{\mathrm{z}i}\right]^{1/2}, \\
    \sigma_\mathrm{rms} = \langle \sigma^2\rangle^{1/2} = \left[\frac{1}{n}\sum_{i=1}^{n} \sigma^2_{i}\right]^{1/2}, \\
    \chi^2/\nu = \frac{1}{n}\sum_{i=1}^{n}\left(\frac{B_{\mathrm{z}i}}{\sigma_{i}}\right)^2.
\end{aligned}
\end{equation}

Here, $B_{\mathrm{z}i}$ and $\sigma_{i}$ are the individual $i$th measured value of the longitudinal magnetic field with error, and $n$ is the total number of measurements.

\begin{figure*}
\centering\includegraphics[width=0.8\textwidth]{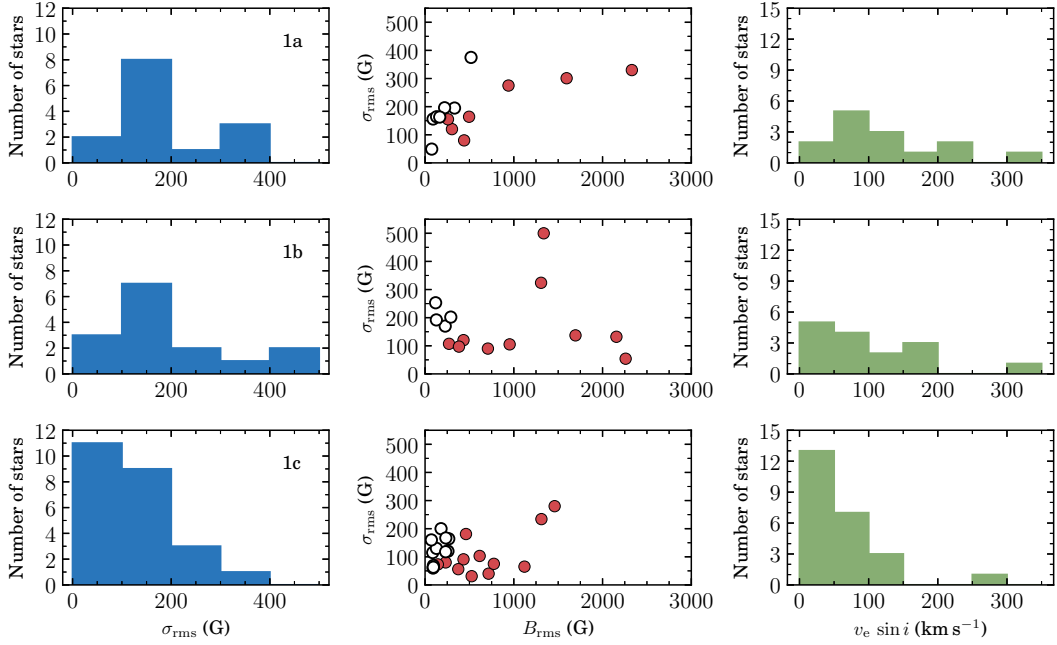}
\caption{Illustration of the measured uncertainties of the longitudinal magnetic field and their correlation with the rotation velocity of the stars in subgroups 1a\:--\:1c. Filled symbols in the middle group of plots mark the stars with detected magnetic field ($\chi^2/\nu \geq 5$).}
\label{fig:detection}
\end{figure*}

We count the stars with $\chi^2/\nu \geq 5$ as magnetic. In Table~\ref{tab:tab1}, 31 stars comply with this criterion. Among them, there are 14 stars whose magnetic fields have been detected in observations with the 6-m SAO telescope. The actual number of the magnetic stars is probably higher, but their fields then must be below the instrumental detection limit achieved in the current work. From the magnetic field measurements of standard stars (Sec. \ref{technique}) under the same signal-to-noise level in a single observation, we estimate this limit (the $3\sigma$ level) as 200\:--\:400\,G for the stars with slow or moderate rotation, and 500\:--\:800\,G for the fast rotators with $v_\mathrm{e}\sin i\gtrsim90$ km\,s$^{-1}$. Although to a less degree, the effective temperature of the stars also affects the accuracy, with growth reducing the number of lines suitable for measurement. For the Orion stars, graphically we show this in Fig. \ref{fig:detection}. Each row of the figure represents one of the three subgroups 1a\:--\:1c. In the left vertical panels of Fig. \ref{fig:detection}, we demonstrate the distribution of $\sigma_\mathrm{rms}$ for each of the three subgroups. The relationship between $\sigma_\mathrm{rms}$ and $B_\mathrm{rms}$ is depicted in the middle vertical panels. The right vertical panels demonstrate the distribution of measured projected rotational velocities $v_\mathrm{e}\sin i$. Comparing the distribution of $\sigma_\mathrm{rms}$ with the velocity $v_\mathrm{e}\sin i$, one can see their strong correlation confining the upper limit of detection.

From the detection statistics, the fraction of magnetic stars among the population of CP stars in Orion OB1 is at least 55\%. Table~\ref{tab:tab2} shows the distribution of these stars between subgroups. For the table, the age $\log t$ of the subgroup was taken from the original paper by \citet{1994A&A...289..101B}. In columns, for each subgroup, through $N_\mathrm{tot}$, $N_\mathrm{CP}$, and $N_\mathrm{mag}$, we give the numbers of early-type members (following \citealt{1994A&A...289..101B}), identified peculiar stars, and confirmed magnetic CP stars, respectively. Then, the fraction of CP stars in the whole population is $f_\mathrm{CP-tot}=N_\mathrm{CP}/N_\mathrm{tot}$, $f_\mathrm{mag-CP}=N_\mathrm{mag}/N_\mathrm{CP}$ is a fraction of the confirmed magnetic CP stars, and $f_\mathrm{mag-tot}=N_\mathrm{mag}/N_\mathrm{tot}$ is a fraction of magnetic stars in the whole subgroup population. The Clopper-Pearson intervals for a binomial distribution are specified next to the value.

Looking at the data in Table~\ref{tab:tab2}, we arrive at the first major conclusion of our work: in Orion OB1, the fraction of CP stars sharply decreases with age. The null hypothesis $H_\mathrm{0}$ asserting the equality of values for 1a and two other subgroups after binomial testing \citep{zar2010biostatistical} is rejected at the significance level $\alpha=0.05$. The same hypothesis being verified in the pairs 1b\:--\:1c and 1c\:--\:1d does not pass the test at the level $\alpha\approx0.15$. The occurrence of a magnetic field in the peculiar stars follows the same yet more gradual trend. Here, the statistical testing does not show the meaningful differences between $f_\mathrm{mag-CP}$ in subgroups 1a to 1c. Given the $p$-values 0.06 and 0.12 obtained for pairs 1b\:--\:1c and 1a\:--\:1c, respectively, we do not reject $H_\mathrm{0}$ and consider the result as marginal. Subgroup 1d is a notable exception and will be considered in the next section.

\begin{table}
\setlength{\tabcolsep}{2pt}
\renewcommand{\arraystretch}{1.35}
\caption{Occurrence of CP and magnetic CP stars in different subgroups of the association with corresponding 95\% confidence intervals.}
\label{tab:tab2}
\begin{center}
\begin{tabular}{lccccccl}
\hline
Subgroup   &  $\log t$  &   $N_\mathrm{tot}$  &  $N_\mathrm{CP}$  & $N_\mathrm{mag}$  & $f_\mathrm{CP-tot}$ &  $f_\mathrm{mag-CP}$  &  $f_\mathrm{mag-tot}$ \\
\hline
1a         &       7.05   &   267   &  14  &   7   & $0.05_{\:0.03}^{\:0.09}$    &  $0.5_{\:0.23}^{\:0.77}$    &  $0.026_{\:0.01}^{\:0.05}$ \\
1b         &       6.23   &   127   &  15  &  11   & $0.12_{\:0.07}^{\:0.19}$     &  $0.73_{\:0.45}^{\:0.92}$   &  $0.087_{\:0.04}^{\:0.15}$ \\
1c         &       6.66   &   276   &  24  &  13   & $0.09_{\:0.06}^{\:0.13}$     &  $0.54_{\:0.33}^{\:0.74}$   &  $0.047_{\:0.03}^{\:0.08}$ \\
1d         &      $<6.0$  &   14    &   3  &   0   & $0.21_{\:0.05}^{\:0.51}$     &  0      &  0     \\
\hline
\end{tabular}
\end{center}
\end{table}

\begin{figure}
\centering\includegraphics[width=0.9\columnwidth]{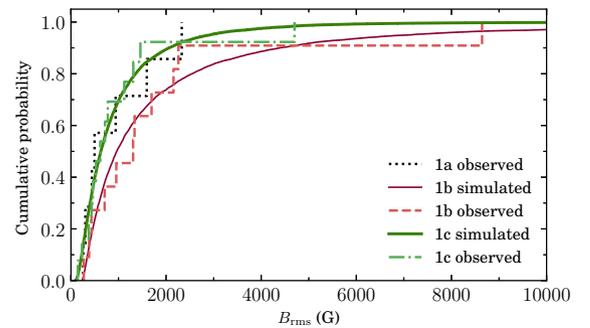}
\caption{Cumulative distribution of the root-mean-squared magnetic field of magnetic stars in subgroups 1b and 1c.}
\label{fig:bzcdf}
\end{figure}

Search for an adequate criterion for characterising the magnetic properties of the entire subgroup is a complex problem. In this paper, we rely on the frequently used criterion~--- the root-mean-squared magnetic field~--- taken for the magnetic members of the corresponding subgroup only. The mean values of $B_\mathrm{rms}$ are listed in Table~\ref{tab:tab3}. In Fig. \ref{fig:bzcdf}, we show the cumulative distribution of $B_\mathrm{rms}$ for the magnetic members in three subgroups.

\begin{table}
\caption{The average root-mean-squared magnetic field $B_\mathrm{rms}$ of different subgroups of the association. Corresponding values for non-detections are referenced in parentheses.}
\label{tab:tab3}
\begin{center}
\begin{tabular}{lcccl}
\hline
Subgroup &  $\log t$ &  $B_\mathrm{rms}$ (G)  &   $\sigma_\mathrm{rms}$ (G)  &  $\chi^{2}/\nu$ \\
\hline
1a      &    7.05   &    1162 (263)    &    223 (206)   &    20  (1.5)   \\
1b      &    6.23   &    2902 (204)    &    252 (207)   &    243 (1.1)   \\
1c      &    6.66   &    1508 (174)    &    160 (132)   &    106 (2.0)   \\
\hline
\end{tabular}
\end{center}
\end{table}

One can note that the subgroups 1a and 1c are generally characterised by the same field $B_\mathrm{rms}$, however the confidence level is much higher for more populated subgroup 1c. The strongest magnetic field is observed in the youngest among the three subgroups 1b, and it is almost three times higher than the one for the oldest subgroup 1a.

To evaluate the statistical significance of the result, we used the Mann-Whitney $U$ test, based on comparison of mean ranks and aimed to show the presence of a shift between the two distributions. The cumulative distribution function (CDF) of the measured magnetic fields (Fig. \ref{fig:bzcdf}) was used to check the null hypothesis suggesting that the CDF for subgroup 1b is stochastically smaller (shifted towards the larger values of $B_\mathrm{rms}$) compare to the corresponding distributions for two other samples.

Pairwise comparison between different subgroups shows that the root-mean-squared magnetic field of subgroup 1b is stronger than the ones of 1c ($p=0.90$) and 1a ($p=0.81$). The same test applied to subgroups 1a and 1c demonstrates no difference between them ($p<0.01$). Due to the statistical similarity of these CDFs, we excluded subgroup 1a from further analysis. To reproduce the underlying lognormal distributions for subgroups 1b and 1c, we estimated the initial parameters from the empirical data and ran two Monte Carlo simulations with 1\,000 trials each. The results of the simulation are plotted as thin solid lines in Fig. \ref{fig:bzcdf}. In the next step, we applied the Kolmogorov--Smirnov goodness-of-fit test (KS) to the obtained curves. The corresponding $p$-values of KS test are 0.86 for subgroup 1b and 0.96 for 1c. The $U$ test confirms that the simulated distributions are distinctively different with high confidence ($p>0.99$). Thus, we come to the second conclusion of our research that the stars observed in subgroup 1b have statistically stronger magnetic fields.

\begin{figure}
\centering\includegraphics[width=0.9\columnwidth]{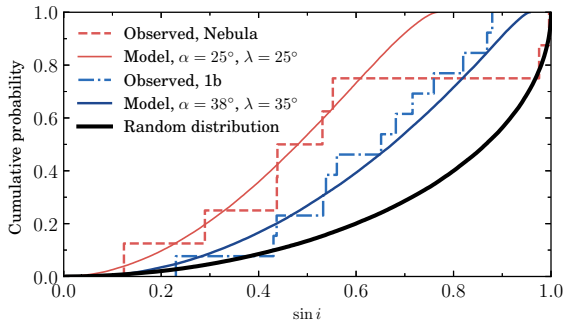}
\caption{The probability distribution of $\sin i$ of the Nebula and the subgroup 1b CP stars. Thin solid lines fit the empirical distributions of spin axes \citep[see][for details]{2010MNRAS.402.1380J}. Black solid line corresponds to the random spin-axis distribution.}
\label{fig:align}
\end{figure}

\subsection{Stars in the Orion Nebula}

Having analyzed the data described in Section~\ref{onc}, we noticed a low incidence of magnetic CP stars in the region of the Orion Nebula. In subgroup 1c, 12 of 25 stars are located inside the nebula. A strong magnetic field was detected in 3 of these stars. All the stars of subgroup 1d are located within the nebula, but none of them show a significant field. Since there are only 13 magnetic stars in subgroup 1c, it turns out that outside of the nebula, we have 9 magnetic CP stars and only 2 non-magnetic ones. As a result, we obtain a contrasting effect: the fraction of magnetic CP stars outside the Orion Nebula is 82\% vs only 20\% of stars inside it.

Now compare the values of the mean longitudinal magnetic field of the stars inside and outside the nebula. For 3 inner magnetic stars, we get $B_\mathrm{rms} = 964\pm171$\,G and $\chi^2/n = 73.5$. In contrast, the corresponding values for 10 external stars are much higher: $B_\mathrm{rms} = 1641\pm158$\,G and $\chi^2/n = 498.5$. Thus, in subgroup 1c, the average magnetic field of stars is almost twice stronger than in the nebula, and this is the third major conclusion of our work.

\section{Discussion}
\label{discus}

This paper summarises the results of a spectropolarimetric survey of the magnetic chemically peculiar stars in the Orion OB1 association. Out of 85 initially selected CP stars in four subgroups of the association, we observed 56 potentially magnetic targets using the Main Stellar Spectrograph of the 6-m telescope BTA of the Special Astrophysical Observatory. In 2013\:--\:2021, we collected more than 600 circularly polarised spectra, which were used to measure stellar longitudinal magnetic fields in a uniform way.

\subsection*{Magnetism of the Orion stars}
We have found 31 CP stars with a measurable magnetic field which is 55\% of the whole sample. The magnetic field of 14 stars was detected for the first time. In our study, the fraction of peculiar stars aged 2\:--\:10 Myr decreases from 0.12 for the youngest subgroup 1b to 0.05 for the oldest subgroup 1a. The occurrence of the magnetic field follows the same, though less articulated, trend. Surprisingly, we have not detected any strong magnetic fields in stars around the centres of star formation regions. Inside the Orion Nebula, there is only one star, HD\,37017, with the longitudinal field exceeding 2\,kG. We suggest that it results from a very young age of stars inside the nebula and, as a consequence, the lower incidence of magnetic field in this type of objects (7\% in \citealt{2013MNRAS.429.1001A, 2013MNRAS.429.1027A}). Besides, the spectral material obtained with MSS during the observational phase of the project does not allow us to confidently distinguish between MS and pre-MS magnetic stars.

Our conclusion about the occurrence of the magnetic CP stars in Orion OB1 is based on an admittedly small sample. However, given the fact that the number of CP stars even in the most populated open clusters is much smaller, stellar associations suit well for studying the role of the magnetic field in stellar formation and evolution. This is especially true for Orion OB1, where even the faintest CP stars are reachable for accumulation of a rather high signal-to-noise ratio with modern spectropolarimeters.

\subsection*{CP stars of Orion OB1 in a new light}
Recent studies of the 3D structure of the association based on modern satellite sky surveys unveil its complexity with numerous less dependent fragments of different ages \citep{2019A&A...631A.166K, 2020A&A...643A.114C, 2020A&A...643A.151R}. This fact forced researchers to reconsider existing scenarios of sequential stellar formation in Orion. It is obvious now that Blaauw's approach was just a very simplified description of the apparent structure and evolution of Orion OB1. In terms of the elucidation of the origin of hot magnetic stars, this new fragmented picture of the association requires reconsidering the stellar ages and initial conditions in the birthplaces of the CP stars being studied. The aforementioned studies lack of useful data about any meaningful fraction of our sample. For those stars which were identified in the literature, we see an inappropriate determination of stellar parameters, primarily, the effective temperatures, radial velocities and age. 
Thus, despite the low density of CP stars in different subgroups, we still consider the ages of selected stars approximately equal to the average age of corresponding subgroups as published in \cite{1994A&A...289..101B}. The age, different from what is stated here, may question the correctness of our interpretation of the evolutionary effects in the occurrence of  CP stars in different subgroups. In the case of more uniform ages, the discovered effects might result from the location-specific stellar formation.

Hereafter we are considering two possible explanations for the fact that different spatially and kinematically separated subgroups of the association demonstrate different magnetic properties. The first interpretation relies on the peculiar rotation of the stars in a selected volume, e.g. the predominant alignment of rotational axes, or systematically different rotation of the whole sample. The second explanation is based on a specific evolution of fossil magnetic fields in hot stars. In the next paragraphs, we consider both in detail.

\subsection*{Stellar rotation}
The ordered rotation of an isolated group of stars, within the frame of the oblique-rotator model \citep{1950MNRAS.110..395S}, can produce systematically biased measured magnetic characteristics. Collimation of the rotational axes is unlikely possible in sparse stellar structures, but in Orion OB1, there are at least two relatively compact groups where the rotational effects can occur. To examine this problem, we analysed the rotational properties of the stars in two regions comprising the Orion Belt (1b) and the Orion Nebula (1c and 1d). For the selected stars, we calculated $\sin i$ as $$\sin i = \frac{P_\mathrm{rot}\,v_\mathrm{e}\sin i}{50.6\,R},$$ 
with the period $P_\mathrm{rot}$ and projected rotational velocity $v_\mathrm{e}\sin i$ taken from Table \ref{tab:tab1}, and stellar radii expressed in solar units with respect to their spectral type\footnote{\url{http://www.pas.rochester.edu/\~emamajek/EEM_dwarf_UBVIJHK_colors_Teff.txt}}.

We successfully evaluated $\sin i$ for 13 CP stars in the subgroup 1b and for 8~--- in the Orion Nebula. For several slow-rotating stars  in our list, the measured $v_\mathrm{e}\sin i$ is limited by the resolving power of the spectrograph. These stars were excluded from further consideration despite the known periods of rotation.

We studied the distribution of $\sin i$ within the frame of a model described by \cite{2010MNRAS.402.1380J}. In this model, the spin vectors of individual stars are distributed within a cone with an opening half-angle $\lambda$ and inclined to the line of sight by the angle $\alpha$. In these terms, $\lambda$ being close to 90$^\circ$ corresponds to a completely random distribution of the spin axes. The smaller the half-angle, the higher is the order of the axis alignment. Eventually, neither of two our samples showed the uniform distribution. For 13 stars of subgroup 1b, the best fit was achieved with $\alpha\approx38^\circ$ and $\lambda\approx35^\circ$. The sample of 8 stars in the Nebula produces the distribution fitted by the model with $\alpha\approx24^\circ$ and $\lambda\approx26^\circ$ (Fig. \ref{fig:align}). Two-sample KS test confirms that the inclination angles in both samples may originate from the same distribution, while this distribution is distinctively not random with a high ($>95$\,\%) level of confidence. We explain this discrepancy by incomplete accounting of slow rotators and imprecise values of stellar radii. Precise values of radii have crucial importance for such a test and we plan to evaluate them in the future works. At this step, we can conclude that the consistent orientation of spin vectors in two compact regions of the association cannot be responsible for the observed differences of the magnetic field.

\begin{figure}
\centering\includegraphics[width=0.9\columnwidth]{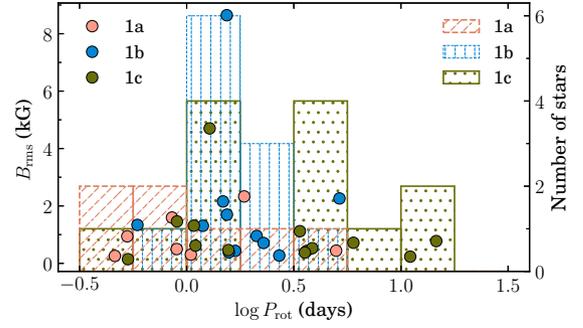}
\caption{The distribution of the rotational periods in different subgroups in Orion OB1 vs the root-mean-square magnetic fields.}
\label{fig:rotfield}
\end{figure}

To explore the possible impact of the stellar rotation on the magnetic field strength, we searched for a possible correlation between the rotation rate and the root-mean-squared field of the stars. Even a short glance at Fig. \ref{fig:rotfield} reveals no significant correlation between $P_\mathrm{rot}$ and $B_\mathrm{rms}$. More accurate analysis shows a moderate correlation between the two parameters with the Spearman coefficient $r=0.21$ for subgroup 1a, $-0.33$ for 1b, and $-0.20$ for 1c. In terms of stellar rotation, the oldest subgroup 1a shows an excess of fast-rotating stars, while the slow rotators occur solely in 1c.

\begin{figure}
\centering\includegraphics[width=0.9\columnwidth]{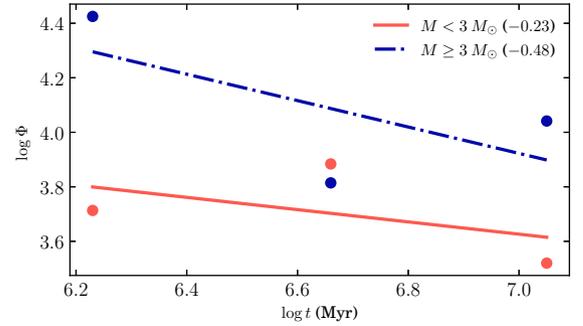}
\caption{Dependence of the magnetic flux $\Phi=B_\mathrm{rms}\,(R/R_\odot)^2$ from time for stars of different masses. The slope is specified in parentheses.}
\label{fig:decay}
\end{figure}

\subsection*{Evolution of the magnetic field strength}
Our survey of the magnetic field in the Orion CP stars shows that the field $B_\mathrm{rms}$ in the young subgroup 1b is almost three times stronger than that in older subgroups 1a and 1c~(Table \ref{tab:tab3}). This fact refers us to the study by \citet{1981ApJ...249L..39B}, who, on a sample of 13 Orion stars, concluded that the field of the younger stars is stronger than that of the older population by a factor of three. From known stellar ages, \citet{1981ApJ...249L..39B} estimated a time scale of the complete decay of magnetic fields to 100 Myr. Nothing similar was observed in the association Scorpius--Centaurus, a yet closer stellar complex containing a significant number of magnetic stars \citep{1987ApJS...64..219T}.

\begin{figure*}
\centering\includegraphics[width=0.85\textwidth]{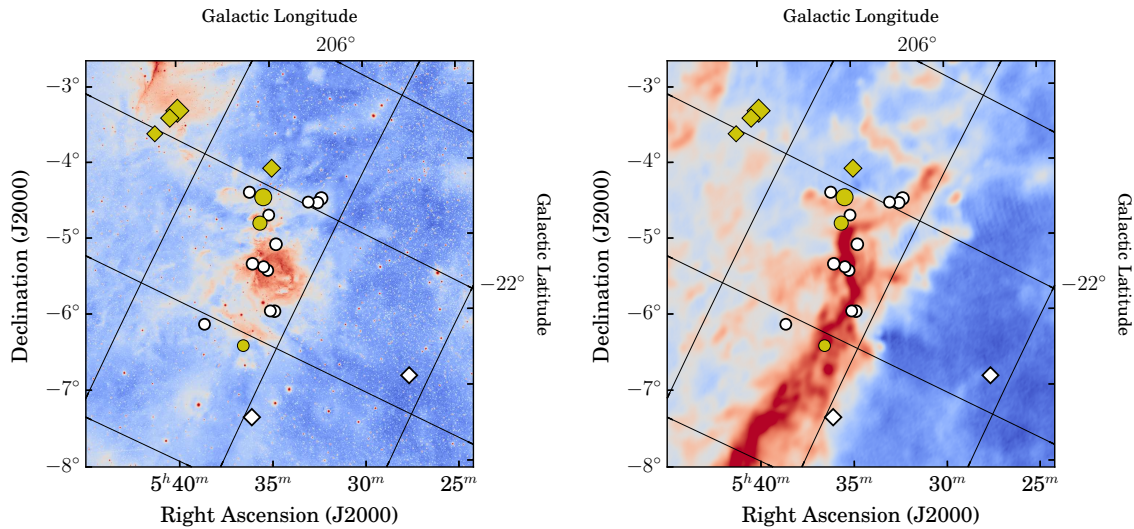}
\caption{Zoomed part of Fig. \ref{fig:scheme} showing the region of the Orion Nebula and its gas and dust complexes.}
\label{fig:closeup}
\end{figure*}

The magnetic field strength decaying with age naturally follows from the theory of the fossil field evolution, though its rate is still a matter of debate. On large samples of heterogeneous observations carried out in different years, \cite{2006A&A...450..763K}, \cite{2007A&A...470..685L}, \cite{2019MNRAS.483.3127S}, and \cite{2019MNRAS.490..274S} reported that the magnetic field of CP stars of different mass declining with age. The authors found that the decay rate correlates with stellar mass and tends to increase for the most massive magnetic stars. To evaluate this rate from our data, we explored the relationship between the unsigned magnetic flux $\Phi=B_\mathrm{rms}\,(R/R_\odot)^2$ and age $\log t$ for the low- and high-mass stars with respect to the stellar mass value $3\,M_\odot$. We used the same source of the stellar radii and masses as for the study of the rotational properties described above. We preferred the magnetic flux to $B_\mathrm{rms}$ to reduce the impact of evolutionary effects caused by increasing stellar radii. As a result, we found that in 17 more massive stars of our sample, the flux decays at a rate $\mathrm{d}\!\log\Phi/\mathrm{d}\!\log t=-0.48$ which virtually coincides with the value $-0.5$ reported in \cite{2019MNRAS.490..274S}. With a slightly lower probability, we also observe the decreasing trend in data for 14 less massive stars, while for this range of masses, the above-cited authors find the flux to be constant with time. Going further, we analysed fluxes of the 8 most massive stars with $M\geq4\,M_\odot$. This analysis gives the slope $-0.63$, confirming faster flux decay with increased mass, in agreement with the above-cited papers. In this way, on a highly uniform sample of young magnetic stars observed within a very limited volume of the Galaxy, we confirm the previous reports of decaying magnetic fluxes in massive CP stars derived from less uniform than in our study samples.

\subsection*{CP stars and interstellar medium}
In contrast to the open clusters and similar compact structures, the association in Orion offers a unique chance to explore the role of a magnetic field in the processes of stellar formation and evolution. Due to the young age of the stellar population, its specific orientation and spatial motion, in Orion, we can trace the individual stars to their suspected birthplaces.

Orion is rich in gas and dust complexes. For the purposes of our study, the most important ones are those comprising chemically peculiar stars. Figure \ref{fig:scheme} demonstrates the distribution of identified CP stars in Orion. Simple visual analysis showed that some stars tend to appear in the zones of increased emission. The region of the Orion Nebula integrated into a northern part of the Orion A molecular cloud is of particular interest. A close-up image of this region in the Planck data (right panel of Fig. \ref{fig:closeup}) reveals that the positions of chemically peculiar stars strictly follow the main filament of the cloud. Even if we consider the complex internal structure of Orion A and its inclination to the line of sight, this pattern does not look like a simple accident.

In \cite{2020A&A...643A.151R} it has been shown that Orion A forms a 3D structure extending to up 500 pc. Towards the heading northmost part of Orion A, the authors have found a foreground cloud of dust containing very few young and many older stellar objects. The authors believe that this cloud can be associated with the new stellar groups identified, e.g., in \cite{2018AJ....156...84K} and \cite{2019A&A...628A.123Z}. These groups are somewhat older than the youngest stars in the Orion Nebula Cluster, with an estimated age of 5\:--\:10 Myr. For us, it means that at least several stars shown in Fig. \ref{fig:criterion:distance} might be older than is adopted in this paper. Nevertheless, assuming the foreground cloud is a remnant of a previous episode of star formation, it can be considered the older integral part of a bigger complex.

The role of the magnetic field in the early stages of star formation is still far from being completely understood. The up-to-date review on this of this problem recently has been given by \cite{2022arXiv220311179P}. Within the scope of this work, 3D mapping of large-scale magnetic field in Orion is especially important. According to \cite{2022A&A...660L...7T}, in Orion A, the field forms a tilted semi-convex structure. The authors found that the strength of the line-of-sight field component increases towards the north side of the Galaxy. Moreover, the field of the cloud in general retains a memory of the initial structure of the Galactic magnetic field. In relation to the considered region of the association, it means that all peculiar stars linked to Orion A and substructures had the same or almost the same initial conditions, including the primordial magnetic field. Thus, the observed characteristics of the CP stars developed later during the individual evolution and vary from star to star. In this light, it appears very promising to study the physical and rotational properties of hot stars in the Orion Nebula and around it. At the moment, the most critical factor impeding such the studies is the lack of measured radial velocities that turns into a large number of hidden spectroscopic binary systems in Orion. We expect to draw a realistic picture of the magnetic and chemical anomalies and their evolution in Orion OB1 after finishing the study of all individual chemically peculiar members.

\section*{Acknowledgements}
The authors thank the anonymous reviewer for their careful reading of our manuscript and their insightful comments and suggestions.

This study was supported by the Russian Science Foundation (RSF grant 21-12-00147). The observations were carried out using the 6-m telescope BTA of the Special Astrophysical Observatory. In this study, the authors used the data publicly available in NASA ADS, SIMBAD, and VIZIER databases.

IY is grateful to RFBR (grant 19-32-60007) for the financial support.


\section*{Data Availability}

The raw observational material in the form of 2D CCD spectra is available through the observational archive of SAO: https://www.sao.ru/oasis/cgi-bin/fetch?lang=en. The authors can provide the extracted 1D spectra upon request.




\bibliographystyle{mnras}
\bibliography{orion_paper}







\bsp	
\label{lastpage}
\end{document}